\documentclass[12pt]{article}
\usepackage{epsfig,amssymb,latexsym,amsmath}
\usepackage{amssymb}  

\newcommand{\be}{\begin{equation}}
\newcommand{\ee}{\end{equation}}

\usepackage{graphicx}

\title{Reality in quantum mechanics, \\
Extended Everett Concept, and consciousness}
\author{Michael B. Mensky\\
{\small P.N.Lebedev Physical Institute,} 
{\small 53 Leninsky prosp., 119991 Moscow, Russia}}
\date{}
 
\begin{document}

\maketitle

\begin{abstract}

Conceptual problems in quantum mechanics result from the specific quantum concept of reality and require, for their solution, including the observer's consciousness into quantum theory of measurements. Most naturally this is achieved in the framework of Everett's ``many-worlds interpretation'' of quantum mechanics. According to this interpretation, various classical alternatives are perceived by consciousness separately from each other. In the Extended Everett Concept (EEC) proposed by the present author, the separation of the alternatives is {\em identified} with the phenomenon of consciousness. This explains classical character of the alternatives and unusual manifestations of consciousness arising ``at the edge of consciousness'' (i.e. in sleep or trance) when its access to ``other alternative classical realities'' (other Everett's worlds) becomes feasible. Because of reversibility of quantum evolution in EEC, all time moments in the quantum world are equivalent while the impression of flow of time appears only in consciousness. If it is assumed that consciousness may influence onto probabilities of alternatives (which is consistent in case of infinitely many Everett's worlds), EEC explains free will, ``probabilistic miracles'' (observing low-probability events) and decreasing entropy in the sphere of life. 

\end{abstract}


\newpage
\section{Introduction}
\label{Sec:Intro}

Paradoxes of quantum mechanics and the resulting so-called ``problem of measurement'' are known from the early years of   quantum mechanics, but are not finally resolved up to now. An essential step in the attempts to solve these problems was made by Everett in its famous ``many-worlds'' interpretation of quantum mechanics \cite{Everett57,DeWittGrah73everett}. In our days the Everett's approach became much more popular. One of the reasons is that it may in a sense be connected with the problem of consciousness (see, e.g., \cite{Squires94bk,Lockwood96mind,Whitaker00,Zeh00,Stapp01cons}). Here we shall discuss the approach called Extended Everett Concept (EEC) suggested by the author \cite{Men00consEn,Men04consPhilosEn,Men05consEn,MenBk05consEn}. This approach allows one to introduce the connection between quantum mechanics and consciousness in a very natural way. The resulting advantage is that some features of consciousness as well as some known but not yet explained phenomena of life directly follow from EEC. Moreover, it directly follows from EEC that these phenomena appear in a special state of consciousness which may be described as being ``at the edge of consciousness'' that may be identified as sleep or trance. 

In Sect.~\ref{Sec:Paradoxes} we shall very briefly show how conceptual problems of quantum mechanics follow from the contradiction between its linearity and the postulate of reduction in the description of quantum measurements. In Sect.~\ref{Sec:Everett} the interpretation of quantum mechanics suggested by Everett for overcoming this contradiction will be presented as well as its extension (EEC) leading to the quantum definition of consciousness. Finally in Sect.~\ref{Sec:EdgeCons} important consequences of EEC (such as the explanation of the phenomenon of life, free will and permanent support of health in an organism) will be reviewed. In Sect.~\ref{Sec:Conclus} a short conclusion will be given.

\section{Quantum measurements: theory and paradoxes}
\label{Sec:Paradoxes}

There is no need to discuss the conceptual problems (paradoxes) of quantum mechanics in detail because they are well known. Let us only mention that all of them follow from special features of the concept of reality in quantum mechanics. These features were first explicitly formulated in the paper by A.~Einstein, B.~Podolsky and N.~Rosen \cite{EinstPodolRosen35}, reformulated later in a more convenient form by John Bell \cite{Bell64,BellBk87}, and experimentally confirmed in the experiments of A.Aspect \cite{Aspect81,Aspect82}. 

The essential difference of the quantum-mechanical concept of reality from usual classical reality is that in quantum mechanics the properties of material systems, as they are observed in a measurement, may not exist before the observation (measurement). If for example the measurement shows that a particle is located in one of two points $A_1$, $A_2$, this particle may be located neither in $A_1$, nor in $A_2$ before the measurement. This is the case if the state of the particle before the measurement, $\psi=c_1\psi_1+c_2\psi_2$, is a superposition of the states $\psi_1$, $\psi_2$ localized correspondingly in $A_1$ and $A_2$. 

According to von Neumann reduction postulate, after the measurement distinguishing between these two alternatives, the system having been previously in the state $\psi$ goes over into one of the states $\psi_1$ and $\psi_2$, with the corresponding probabilities $|c_1|^2$ and $|c_2|^2$. This postulate corresponds to what is observed in real measurements, so the reduction postulate is accepted as the basis for the quantum-mechanical calculations. However, it contradicts to the linearity of quantum mechanics when the process of measurement is considered as an interaction of two systems (the measured system and the measuring device). 

Let the initial state of the device be $\Phi_0$ and the initial state of both systems, $\psi_i\Phi_0$, goes over, after the interaction described by the unitary evolution operator $U$, into $U\psi_i\Phi_0=\psi_i\Phi_i$. Then it follows from the linearity of the operator $U$ that the initial state $\psi\Phi_0$ changes, in the course of the interaction, as follows:  
$$ 
\psi\Phi_0\rightarrow U\psi\Phi_0=U(c_1\psi_1+c_2\psi_2)\Phi_0
=U(c_1\psi_1\Phi_0+c_2\psi_2\Phi_0)=c_1\psi_1\Phi_1+c_2\psi_2\Phi_2.
$$
If one include in the description not only the measuring device, but also the observer as one more physical system initially in the state $\chi_0$, and apply usual quantum-mechanical consideration to the three systems, then their evolution under the interaction will be given as follows: 
\be \label{SuperposeObserv} 
\psi\Phi_0\chi_0=(c_1\psi_1+c_2\psi_2)\Phi_0\chi_0 \rightarrow c_1\psi_1\Phi_1\chi_1+c_2\psi_2\Phi_2\chi_2.
\ee

Thus, the linearity of the quantum-mechanical evolution requires that both alternatives 1 and 2 forming the initial superposition $\psi$ exist also (in the superposition with the same coefficients) after the interaction with the measuring device. However, the description of the observation seems to require the reduction, i.e. surviving only a single alternative. The same is of course valid in case when many alternatives are distinguished by the measurement, instead of two of them. The contradiction arises between the linearity of quantum mechanics and the picture of reduction presenting the observation. This contradiction is actually the reason of the quantum-mechanical paradoxes, or conceptual problems. 

We see also that the problem is not overcome by the observer as a physical body being included in the description of the measurement. The key role is therefore played not by the physical body of an observer but by her consciousness. 

\section{Everett (``many-worlds'') interpretation and its extension}
\label{Sec:Everett}

Everett's interpretation is sometimes estimated to be logically complicated. However, it seems complicated only from the point of view of macroscopic, and therefore classical, picture of what happens in the measurement. From the point of view of quantum mechanics the Everett's interpretation is quite simple. Indeed, it excludes the reduction postulate and recovers linear character of quantum mechanics in full volume. Thus the paradoxical character of quantum mechanics is overcome not by inclusion new elements in the theory (and therefore making it more complicated) but by exclusion most unnatural elements of this theory. 

The Everett's interpretation is not so simple in its treating the picture arising before the eyes (in the consciousness) of the observer. We shall see however that this may be essentially simplified in the framework of the Extended Everett Concept (EEC). 

\subsection{Everett interpretation: taking quantum mechanics seriously}

The logic of Everett's interpretation is very simple. We know that the evolution is linear in quantum mechanics. A measurement is nothing else than an interaction between the measured system and its environment (including the measuring device and the observer). Let us take these facts seriously and accept that the measurement is actually linear process. Then the state after the measurement has the form of a superposition, as in Eq.~(\ref{SuperposeObserv}) or in the corresponding formula with summation over many alternatives:
\be \label{SuperposeObservSum}
\psi\Phi_0\chi_0=\sum_i c_i\psi_i \Phi_0\chi_0 \rightarrow \sum_i c_i\psi_i \Phi_i\chi_i.
\ee

Assuming the linearity, Everett must then somehow interpret all terms of the superposition in the right-hand-side of this equation, i.e. all alternative readouts of the measurement. All alternatives are in his concept equally real and should be considered on equal foot. How it may then occur that the observer perceives only one of these alternatives? The answer may also be read off from Eq.~(\ref{SuperposeObservSum}). This formula means that the state of the observer is described also by the various components included in the superposition. The component $\chi_i$ describes the state of the observer in which she sees that the measuring device is in the state $\Phi_i$ thus pointing out that the system is in the state $\psi_i$. This is the picture of a single `Everett's world'. The triplets $\psi_i \Phi_i\chi_i$, with all possible $i$, coexist, forming the ``Everett worlds'', or, more precisely, alternative `classical projections' of a single quantum world. All such ``worlds'', or alternative classical realities, coexist and are equally real (should be considered on equal foot). 

\subsection{Extended Everett concept: quantum consciousness and life}
\label{Ssc:QuConsLife}

It is convenient for us to express the situation in the Everett's interpretation as follows. All classical alternatives are perceived by consciousness (of the observer), however {\em the alternatives are separated by the consciousness}: each alternative is perceived independently from the others. 

Note that all the observers in a single Everett's world (in the same classical reality) see the same, their observations are in complete agreement with each other. This follows from the fact that the initial state with two observers, $\sum_i c_i\psi_i \Phi_0\chi^{(1)}_0\chi^{(2)}_0$ will go over, after the measurement, into the state $\sum_i c_i\psi_i \Phi_i\chi^{(1)}_i\chi^{(2)}_i$ (the crossing terms with $\chi^{(1)}_i\chi^{(2)}_j$, $i\neq j$, cannot emerge). 

We may now concentrate on the whole component of the superposition, $\Psi_i=\psi_i \Phi_i\chi_i$ rather than on its factors $\psi_i$, $\Phi_i$ and $\chi_i$. The right-hand-side of Eq.~(\ref{SuperposeObservSum}) takes then the form 
\be \label{SuperposeClassAltern}
\Psi = \sum_i c_i\Psi_i.
\ee
This equation presents a state of the whole quantum world as a superposition of classical (more precisely, close to classical, or quasiclassical) states of this world. In the previously introduced terms, the whole world contains both the measured system and its environment, including the observer. Now, to move further, we do not need the picture of measurement in these details. Instead, we may talk about the state of the (quantum) world as it is reflected in consciousness, hence the superposition of (quasi)classical states of the world. 

Now we have to do a decisive step, leading to the radical simplification of the whole concept and to very interesting consequences. Taking into account that nobody knows actually what is consciousness, we assume that {\em consciousness is nothing else than the separation of the alternatives}. This identification of consciousness with the separation of the alternatives is a crucial point of the Extended Everett Concept developed in \cite{Men00consEn,Men04consPhilosEn,Men05consEn,MenBk05consEn}. In this assumption, two unclear concepts, one from quantum mechanics and the other from psychology, are identified and thus ``explain each other''. The whole concept becomes simpler. More important is that this leads to new and very interesting consequences. 

First of all, this explains why the alternatives in Eq.~(\ref{SuperposeClassAltern}) should be classical. Instead of the vectors $\Psi_i$ (classical alternatives) we could make use of the other vectors (linear combinations of $\Psi_i$) to present the state of the quantum world $\Psi$ as a superposition. Why have we to take those vectors which are close to classical states? Why the alternatives in the description of consciousness are (close to) classical? The answer is almost evident. 

Consciousness is a feature (and the principal feature) of living beings. (Note that here the term ``consciousness'' means the most primitive, or the most deep, level of consciousness, differing perceiving from not perceiving). If the picture of the world as it is appears in consciousness were far from classical, then, due to quantum non-locality, this would be a picture of a world with ``locally unpredictable'' behavior. The future of a restricted region in such a world could depend on events even in very distant regions. No strategy of surviving could be elaborated in such a world for a localized living being. Life (of the form we know) would be impossible. On the contrary, a (close to) classical state of the world is ``locally predictable''. The evolution of a restricted region of such a world essentially depends only on the events in this region or not too far from it. Influence of distant regions is negligible. Strategy of surviving can be elaborated in such a world for a localized living being. Therefore, classicality of the alternatives $\Psi_i$ is a necessary condition for life. The very concept of life naturally arises in this way from EEC. 

\section{At the edge of consciousness}
\label{Sec:EdgeCons}

It is astonishing that EEC leads to very concrete conclusions about some special features and unusual abilities of consciousness and, even more astonishing, to the concrete characterization of the conditions providing these special abilities. Consciousness is predicted to manifest its unusual abilities when it is almost turned off, i.e. is in the state similar to sleeping or trance. The reason is that in this case consciousness, when working with the given classical alternative, may obtain information from the quantum world as a whole, i.e. from ``other classical alternatives''. This conclusion follows from the definition of consciousness accepted in EEC. If, in addition to this definition, the assumption is accepted that consciousness may influence on probabilities of classical alternatives, then EEC leads also to some well known but not yet explained features of living organisms (such as free will). 

\subsection{Information from `other classical realities': Comparison of alternative realities and predictions}
\label{Ssc:InfoQuantWorld}

\begin{figure}[t]
	\begin{center}
		\includegraphics[scale=0.5]{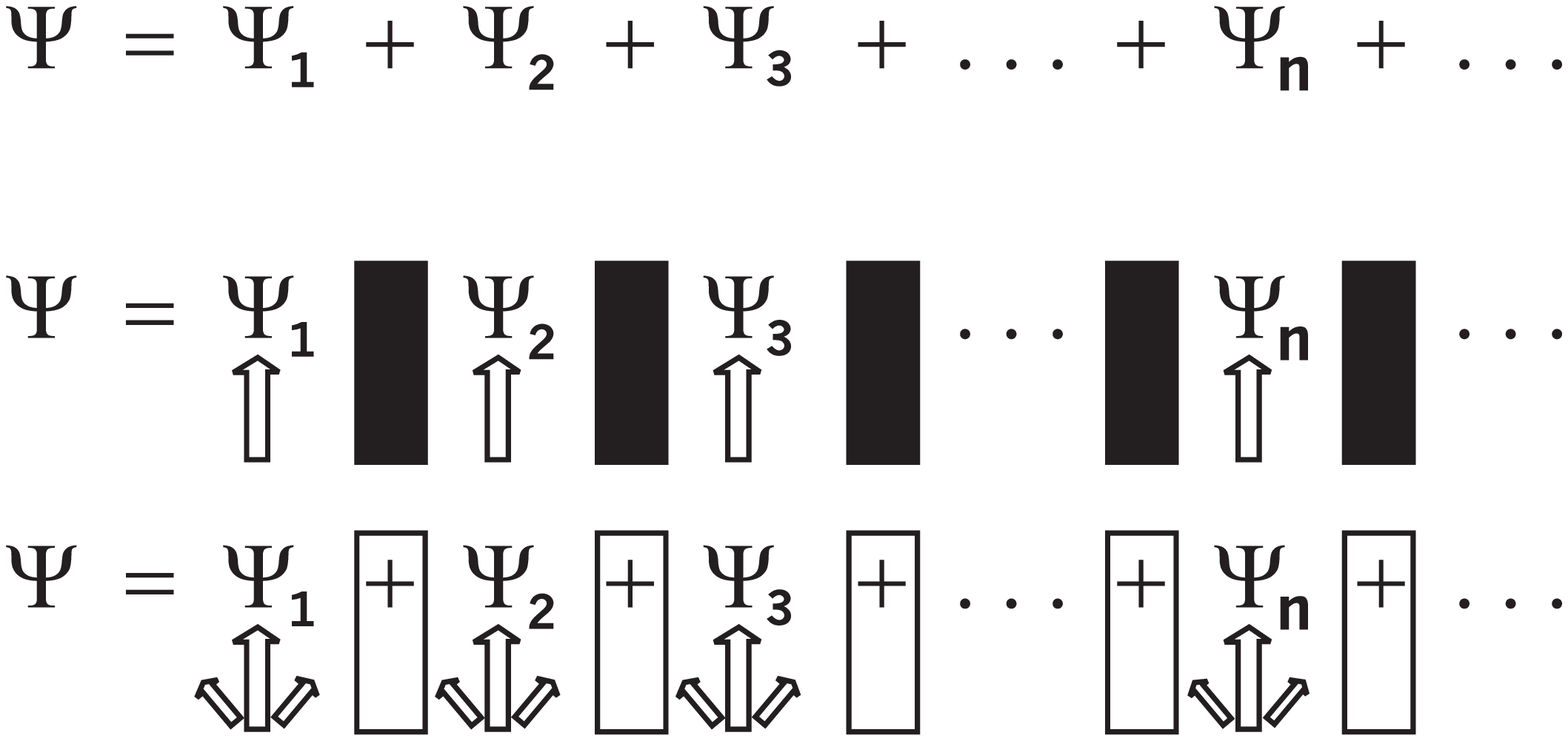}
	\end{center}
	\caption{No separation, or non-alive matter (top diagram); separation, or consciousness (middle); weak separation, or `at the edge of consciousness' (bottom)}
\label{fig:Fig1}
\end{figure}

Let the state of the quantum world be presented by Eq.~(\ref{SuperposeClassAltern}) where each component $\Psi_i$ of the superposition on the right-hand-side is a state presenting a `classical alternative' of the world. Consciousness perceives these alternatives separately from each other. Moreover, according to the definition accepted in EEC, consciousness is {\em identified} with separation of the alternatives. Complete disappearance of consciousness (for example in case of death) means complete disappearance of the separation (just as no separation exist in the description of non-alive matter, with no phenomenon of life). If consciousness does not disappear but becomes weak (in the state of sleep or trance) then, arguing in the same logic, we have to conclude that the separation of alternatives becomes not absolute. The `partitions' between the alternatives become transparent. When perceiving one of the alternative classical realities, the consciousness may then perceive also `other alternative classical realities' (see Fig.~\ref{fig:Fig1}). 
At the moment of returning to the full consciousness (absolute separation) some part of the information from `other realities' may be kept and exploited in the usual work of consciousness with `its own' alternative. 

The same may be formulated in another form: in case of partial turning consciousness off (in the state of sleep or trance) it can extract information from all classical alternatives, or, in other words, from the whole quantum world. At the edge of consciousness one obtains access to the whole quantum world. This should supply additional (as compared with the regular functioning of consciousness) and quite unusual abilities.

What are the features of these additional abilities of consciousness? We can immediately point out two of them. First, information `from other classical alternatives', i.e. from various scenarios possible for a classical world, allows one to compare these scenarios and conclude what scenario is the best (favorable for life). Later (Sect.~\ref{Ssc:FreeWill}) we shall see how this information may be used. 

Second, in the Everett concept the quantum world as a whole (i.e. without separation of the alternatives) is reversible. Its image is a four-dimensional manifold rather than a three-dimensional space developing in time. When consciousness looks out from a single alternative into this reversible world, it can take information from any part of this world. Returning from the quantum world as a whole to `its own classical reality', the consciousness may possess information extracted not only from `other (alternative) scenarios', but also from any stage of these scenarios, including information from the future of each scenario. This argument hints that predictions made in sleep or trance may be possible. We see also that predictions should have relative rather than absolute character. Indeed, they depend on the concrete scenario: the predictions become true only if the given scenario will be realized in the course of the further evolution of `my own classical reality'. 

There are many evidences of successful predictions made by some people in sleep or in trance. Many of these evidences seem to be well documented. The consideration in the framework of EEC may explain both the feasibility of successful predictions and relative character of each prediction, i.e. its not full reliability. One may think that a prediction made in explicitly relative form (something will happen under the condition that something else will do) should be more reliable (if the predictions made by the same person are compared). 

\subsection{Modification of probabilities: Free will and probabilistic miracles}
\label{Ssc:FreeWill}

Up to now our consideration was based only on the Everett's interpretation of quantum mechanics and the identification of consciousness with the separation of alternatives. Let us now accept an additional assumption that {\em consciousness may modify probabilities of classical alternatives}. From the point of view of an observer this means that, perceiving a definite (alternative) classical reality she may have influence on what alternative she will perceive in the next moment (next observation). Probabilities of some of the `next moment' alternatives (which seem to be favorable) may be increased, while the probabilities of others decreased. 

Why this assumption seems to be natural in the context of EEC? It looks natural because in the framework of this concept separation of the alternatives is considered from two qualitatively different points of view. First, from the point of view of quantum mechanics (describing only non-alive matter, including although bodies of living beings when they are considered simply as physical systems), and second, from the point of view of psychology. There is only one universal probability distribution in quantum mechanics ($|c_i|^2$ in the previous example), but the probabilities may in principle be different from the point of view of the consciousness of a living being: various observers may elaborate different probability distributions for what alternatives they are going to see. 

There are two evident objections against this assumption. First one is purely mathematical. The probability of the $i$th alternative is naturally defined as the `relative number' $N_i/N$ of the Everett's worlds of the definite type, such that just the $i$th alternative is realized in all of them. At first glance, this definition is unambiguous and should imply a universal probability distribution. However, this is not the case if the `number of Everett's worlds' is infinite \cite{Men05consEn,MenBk05consEn}. If both $N_i$ and $N$ are infinite, this `definition' becomes ambiguous because of a paradoxical feature of an infinite set: its proper subset may be put in one-to-one correspondence with the whole set. Because of this, different probability distributions on an infinite set are compatible. 

This becomes obvious if one make use of a naive picture where each of the observers sends her `twin-observers', one after another, into various Everett's worlds. Let for simplicity we have two observers and two types of Everett's worlds, $E_1$ and $E_2$ (with the infinite numbers of worlds of each type). One of the two observers may send his `twins' according to the rule: each twin having odd number goes to a world of type $E_1$ and each even twin, to a world of type $E_2$. The other observer may use another rule: each twin with the number divisible by 3 goes to an Everett's world of type $E_1$ while the rest are sent to the worlds of type $E_2$. Then the probability for the first observer to find herself in the world of type $E_1$ is equal to 1/2, while for the second observer it is equal to 1/3. Nevertheless, in case of infinite number of worlds and of twins, all twins will be distributed between the worlds, and each Everett's world will obtain a single twin of the first observer and a single twin of the second one. 

There is another objection against the discussed assumption. If probabilities are not universal, then laws of nature may be violated, but they seem to always endure experimental check-up. The answer to this objection is that the experimental check-up is feasible only for very simple events (such as where an electron should fly etc.). For the events of this type consciousness hardly may modify probabilities because these events are not important from the point of view of living beings. Since laws of physics govern only such simple events (and rather simple combinations of them), the laws of physics should be valid. If the ability of consciousness to modify probabilities may exist, then it should concern only `significant' (for life) events. Such events have much more complicated structure and cannot be reduced to simple events investigated by physicists. Therefore, probabilities of these highly complicated (from the point of view of a physicist) `significant events' cannot be calculated with the help of quantum-mechanical formulas. The question about violation of laws of physics in the scope of such events (in the sphere of life) is therefore meaningless. 

Taking these arguments into account, we may assume that consciousness may modify probabilities of classical alternatives. How this ability of consciousness may manifest itself? First of all it is evident that `probabilistic miracles' become possible under this assumption. This means that consciousness may increase the probability of an event which otherwise seems almost improbable. The probability may even be made close to unity. In the latter case modifying probabilities looks as a choice of a definite alternative. However, this is not a choice but only modification of probabilities, since all non-zero probabilities remain non-zero (although may become very close to zero). The event that is chosen by consciousness (and therefore looks to be a miracle) always has non-zero probability even without modification of probabilities. It is therefore feasible event even in `natural' conditions, without any influence of consciousness. The realization of an event which is characterized by a low probability, does not strictly speaking violate any laws. Instead, it may seem a rare coincidence. For example, if someone says that she wishes to stop rain, and the rain stops, then this may be a probabilistic miracle, but instead it may be interpreted as a coincidence. 

There is one more class of well known and not exotic events which also can be explained with modifying probabilities by consciousness. These are events realizing what is called {\em free will}. If I wish to go to the right and actually go to the right, how this happens? In fact, there is no explanation of this simple ability of consciousness. In the framework of EEC, if the modification of probabilities is assumed, free will is explained quite naturally. There are two alternatives: in one Everett's world I go to the right, in the other one I go to the left. Both alternatives have non-zero probabilities. My consciousness modifies the probabilities increasing the probability of the first alternative. As a result, with a high probability I go to the right. I chose to go to the right. This was my free will. 

It worth noting that in most cases free will is realized `unconsciously', i.e. in the state `on the edge of consciousness' (in this case only a part of the consciousness is `almost turned off', namely the part controlling the body's movements). This is in accord with our prediction that the special abilities of consciousness should manifest themselves just in this state. We shall return to this important point in Sect.~\ref{Ssc:Unconscious}. 

\subsection{Unconscious: the miracle of life}
\label{Ssc:Unconscious}
According to EEC, at the edge of consciousness it has access to information from the whole quantum world, i.e. from various classical realities (classical scenarios). Moreover, if consciousness may modify probabilities of alternatives, it may choose (make more probable) those alternatives (classical realities) that are favorable for life. Thus, the unusual information obtained from the quantum world can be used for improving quality of life. This may shed light on many phenomena in the sphere of life that are well-known and seem quite usual, but in reality have no explanation up to now. These phenomena may in fact be called miracles of life. 

Two important examples are, first, the health and its support and second, the role of sleep. It is commonly believed that health is supported due to the great efficiency of the organism as a self-regulating machine. However, it is difficult to imagine that an organism is efficient enough to support health during the whole life, despite of enormous number of unpredictable damages happening in this life. It seems almost evident that periodic usage of some data base is necessary for correcting these damages. But what is this data base and what is the mechanism of its usage, remains unknown. 

In the phenomenon of sleep, among many astonishing facts, the most strange seems the fact that regular sleeping is absolutely necessary not only for health, but even for very life. A man certainly dies if he is deprived of sleep during few weeks. Why? The common opinion that sleep supplies rest for all systems of organism is evidently insufficient to explain absolute necessity of regular sleeping. 

Our consideration in the framework of EEC suggests an explanation of both these strange features of life: permanent support of health and necessity of sleep. During sleep (or rather during a definite phase of sleep, called paradoxical sleep) a man is at the edge of consciousness and therefore obtains information from alternative classical scenarios. She can compare various scenarios, particularly various scenarios for the body, and find out what scenarios are favorable. Returning, after the sleep, to the usual state, consciousness increases probabilities of just these scenarios. This is a mechanism of permanent support of health. It is known that paradoxical sleep of old people becomes shorter. Perhaps this is the main reason why their health is not supported well enough. 

In this explanation, the hypothetical data base containing recommendations for health is nothing else than the set of all possible scenarios for functioning the body. This data base is always actual because consciousness may compare those scenarios that start from the present state of the body. This returns us to the arguments of Sect.~\ref{Ssc:QuConsLife}. Once more, now on a more concrete level, we may conclude that the mystery and miracle of life is connected with quantum definition of consciousness, as it is given in EEC. 

There is one more unsolved problem in biology that also could obtain its explanation in EEC. This is the problem of morphogenesis. How an embryo is constructed starting from a single cell? Where is a plan of the process of constructing it, step by step, or how constructing is controlled and directed? It is possible that the answer is analogous to the argument above: `consciousness' (the primitive-level consciousness, or ability to somehow perceive, which is connected with a living being from the very beginning) periodically addresses to the quantum world as a whole, compare various scenarios of constructing embryo (various `building plans') and then, returning to the usual state, increase probabilities of those scenarios that lead to the right construction. Of course, this is only a sketch of a possible explanation of the phenomenon, its main idea. 

\subsection{Flow of time and decreasing entropy in the sphere of life}
\label{Ssc:DecreasEntropy}

One of the problems permanently discussed in literature is how irreversibility arises if all equations presenting dynamics of physical systems are reversible in time (see for example \cite{ZehReverse}). In the framework of EEC this problem obtains natural solution: quantum world is reversible (and thus reversible is microscopical theory of non-alive matter, its dynamics), while pictures of (alternative) classical realities appearing in consciousness are irreversible. Indeed, in quantum mechanics irreversibility might appear in the course of reduction, but reduction is excluded from the Everett's interpretation of quantum mechanics (see Sect.~\ref{Sec:Everett}). In the framework of EEC we have, instead of reduction, separation of alternatives which is identified with consciousness. Thus, irreversibility appears only in consciousness. 

The quantum world as a whole, without separation of classical alternatives, is (in EEC) reversible. Its adequate image is given by 4-dimensional manifold with all times considered on equal foot. However, in the picture seen by an observer as an (alternative) classical reality, the `present' time moment is distinguished, radically differs from the past and the future. It is the moment when consciousness chooses (with the help of modification of probabilities) the concrete alternative it will see in the next moment. While the quantum world looks as something given in its wholeness, an (alternative) classical reality is permanently becoming. Time moments in an (alternative) classical reality are divided on the past (when this concrete alternative reality is fixed), the future (when many alternative continuations exist\footnote{Possibility of various continuations of a given `classical reality' seemingly contradicts to the previous statement about `classical world' being locally predictable. However, various continuations are in fact possible (although they cannot differ too much from each other) because the `classical reality' is not precisely classical, but rather close to classical. This possibility may be characterized in an adequate way in terms of continuous quantum measurements (see \cite{Men05consEn}).} for this concrete alternative reality), and the present (when the choice of future is performed). 

One more important unsolved problem is decreasing of entropy in the sphere of life where the processes of self-organization are not only possible but necessary. This contradicts to the general principle accepted in physics and confirmed many times: entropy may only be constant or increase. In the framework of EEC this contradiction disappears because the spheres where entropy correspondingly increases or decreases, are separated. 

The quantum world as a whole is reversible, and its entropy is constant. The entropy of a restricted region of the quantum world may increase. However, in the sphere of life (which includes consciousness working with alternative classical realities) entropy may decrease. Moreover, it is necessary decreasing for the set of classical scenarios realizing life. This is because consciousness regularly compare alternative classical scenarios and choose (by modification of probabilities) those of them which are favorable for life. The deep reason of decreasing entropy is that the concept of the goal arises in the sphere of life. Another important point is that entropy in the sphere of life characterizes not all possible classical scenarios but only those favorable for life (selected according to the goal of improving quality of life). 

\section{Conclusion}
\label{Sec:Conclus}

We have seen that Extended Everett Concept (EEC) may give a quantum definition of consciousness: consciousness is separation of the alternatives. This puts forward a novel view of a difficult question, what is life. Possessing the ability to compare various classical scenarios and choose the most favorable of them, quantum consciousness turns out to be the very essence of life and explains such important phenomena in the sphere of life as free will and efficient support of health. Besides, EEC explains why decreasing entropy in the sphere of life is compatible with the general law of increasing entropy. It becomes clear that entropy in the sphere of life is decreasing because life is presented by a subset of specially selected scenarios. 

\begin{figure}[t]
	\begin{center}
		\includegraphics[scale=0.5]{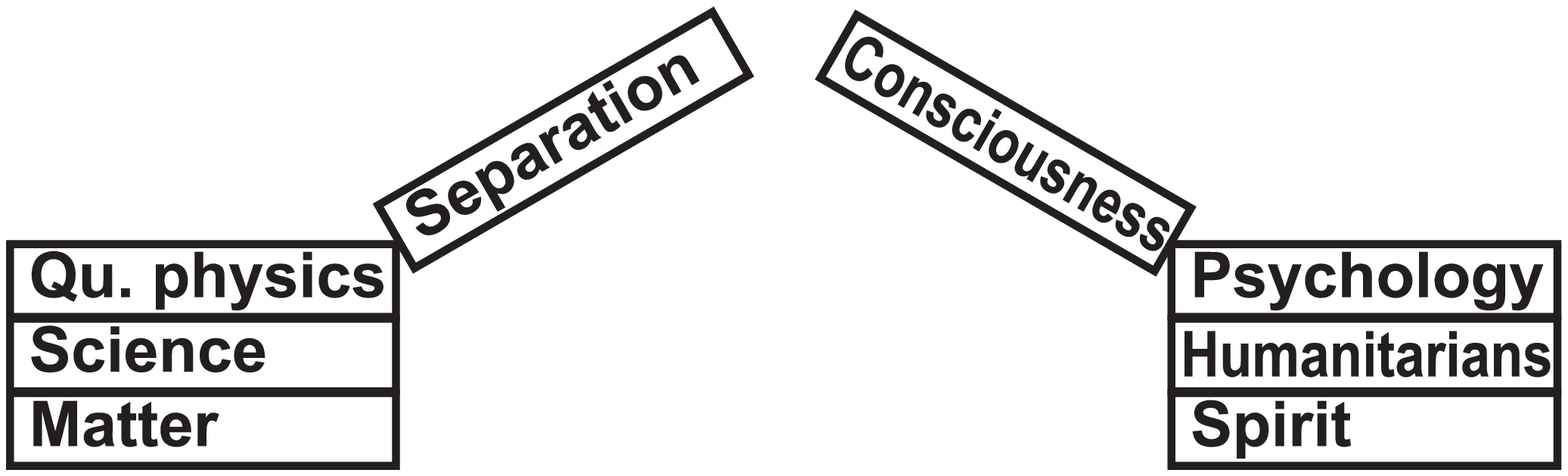}
		
\qquad

	\includegraphics[scale=0.5]{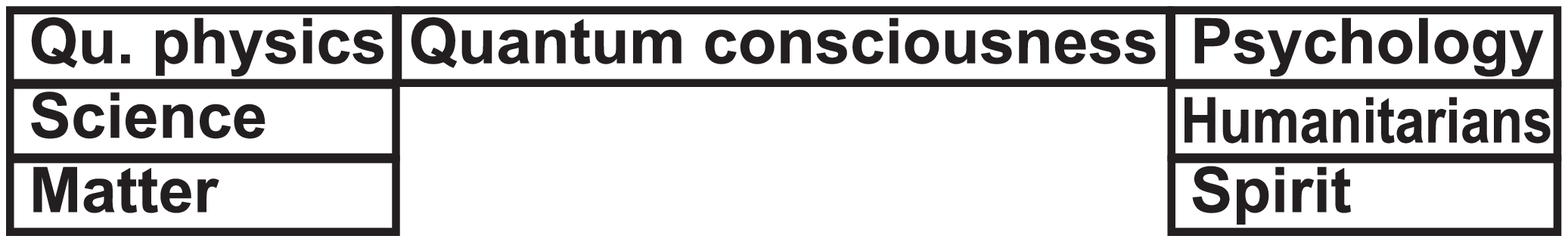}
	\end{center}
	\caption{Lowering the draw-bridge: Identifying `Separation' and `Consciousness' gives `Quantum Consciousness'. This supplies a bridge between natural sciences and humanitarians, between the spheres of matter and spirit}
\label{fig:Fig2}
\end{figure}

Consciousness, as it is defined in EEC, is a general part of two qualitatively different spheres of cognition. Being defined as a separation of alternatives, consciousness is a part of quantum physics, therefore, of natural sciences. Being a special phenomenon characteristic of living beings, consciousness belongs to psychology, therefore, to the humanitarians or, more generally, to the sphere of knowledge about spirit. Thus, quantum consciousness, in the sense of EEC, is a common part of, and provides a bridge between, these two spheres. This seems very interesting because the two spheres are often considered as having nothing in common (although the conceptual relations between them do of course exist and are actively discussed). It seems to us that EEC is one of few approaches that establish deep internal connection between natural sciences and humanitarians, penetrating deeper in the nature of life and human consciousness. EEC lowers the draw-bridge (see Fig.~\ref{fig:Fig2}) 
over the deep precipice dividing two spheres of cognition, the world of matter and the world of spirit. 



\newpage

\end{document}